\documentclass[sigplan,10pt]{acmart}

\newcommand*\rot{\rotatebox{90}}
\usepackage{multirow}
\usepackage{pifont}
\newcommand{\tick}{\ding{59}}

\AtBeginDocument{%
  \providecommand\BibTeX{{%
    \normalfont B\kern-0.5em{\scshape i\kern-0.25em b}\kern-0.8em\TeX}}}

\copyrightyear{2019}
\acmYear{2019}
\setcopyright{acmlicensed}
\acmConference[Middleware 2019]{Middleware 2019: 20th ACM/IFIP International Middleware Conference}{9 - 13 December, 2019}{UC Davis, CA}

\usepackage{tikz}
\usepackage{booktabs}

\usepackage{makecell}

\usepackage{url}
\usepackage{breakurl}
\usepackage{array}
\newcolumntype{L}[1]{>{\raggedright\let\newline\\\arraybackslash\hspace{0pt}}m{#1}}

\DeclareRobustCommand\dash{\textemdash}

\usetikzlibrary{arrows,calc,positioning}
\tikzstyle{intt}=[draw,text centered,minimum size=6em,text width=5.25cm,text height=0.34cm]
\tikzstyle{intl}=[draw,text centered,minimum size=2em,text width=2.75cm,text height=0.34cm]
\tikzstyle{int}=[draw,minimum size=2.5em,text centered,text width=3.5cm]
\tikzstyle{intg}=[draw,minimum size=3em,text centered,text width=6.cm]
\tikzstyle{sum}=[draw,shape=circle,inner sep=2pt,text centered,node distance=3.5cm]
\tikzstyle{summ}=[drawshape=circle,inner sep=4pt,text centered,node distance=3.cm]

\begin{document}

\title[Privacy and Confidentiality on Permissioned DLTs]{Designing for Privacy and Confidentiality on Distributed Ledgers for Enterprise (Industry Track)}

\author{Allison Irvin}
\affiliation{%
  \institution{IBM Research Australia}
}
\email{allison.irvin@au1.ibm.com}

\author{Isabell Kiral}
\authornote{Both authors contributed equally to this research.}
\affiliation{%
  \institution{IBM Research Australia}
}
\email{isabell.kiral@au1.ibm.com}

\begin{abstract}
Distributed ledger technology offers numerous desirable attributes to applications in the enterprise context.
However, with distributed data and decentralized computation on a shared platform, privacy and confidentiality challenges arise.
Any design for an enterprise system needs to carefully cater for use case specific privacy and confidentiality needs.
With the goal to facilitate the design of enterprise solutions, this paper aims to provide a guide to navigate and aid in decisions around common requirements and mechanisms that prevent the leakage of private and confidential information.
To further contextualize key concepts, the design guide is then applied to three enterprise DLT protocols: Hyperledger Fabric, Corda, and Quorum.
%
%
%
%
%

\end{abstract}

\maketitle

\section{Introduction}
Distributed ledger technology (DLT) and blockchain in particular are widely known as the technologies behind digital asset platforms such as Bitcoin \cite{Nakamoto} and Ethereum \cite{Buterin2014}.
These decentralized protocols offer parties the ability to record the exchange of assets and data on an append-only ledger without the involvement of a central authority.
The promise of DLT is to provide a secure and tamper-proof record of every transaction that has ever taken place on a shared network.
Applications of this technology, however, are not limited to the management of assets in the public domain.

In the enterprise context, permissioned DLTs can assist with a number of otherwise hard to manage issues.
Shared ledgers facilitate an audit of past transactions, aid in contract litigation, record consent of parties via digital signatures, and can be used to transparently execute shared and version-controlled business logic.
However, having data replicated across multiple entities inherently poses challenges for the preservation of privacy and confidentiality.
While similar, albeit use case dependent, challenges arise in both public and permissioned ledgers, it can be the legal, regulatory, or contractual requirements that demand particular diligence for the design of enterprise solutions.

In this paper, we will address privacy and confidentiality for (a) the group of interacting parties, (b) transaction data, and (c) business logic.
While public DLTs place a strong focus on pseudonymity, in a permissioned ledger revealing an identity may not only be acceptable but is often a legal requirement.
However, not all businesses may want their relationships with other parties to be visible to unauthorised network members, meaning that enterprise DLT platforms may need to offer the possibility to keep interactions private.
Moreover, data may be strategic to the participating business entities or be sensitive, such as customer data or Personally Identifying Information (PII) (Social Security Numbers, passport details, driver's licenses, etc).
In public blockchains the logic that controls state updates can be fully transparent (as in Bitcoin) or revealed as bytecode (as in Ethereum) with execution of the logic and its results made public.
Meanwhile, for enterprise use cases, any business logic may contain sensitive business information, requiring that code is not shared with all network participants.

In an enterprise solution, the specific use case will determine the privacy and confidentiality requirements of the architecture.
Catering to these requires the architect to navigate a vast space of possible privacy and confidentiality preserving mechanisms.
To facilitate the design of enterprise solutions, the following paper aims to serve as a design guide for systems in the enterprise context that are built on distributed ledger technology.

This paper is structured as follows:
Section \ref{sec:mechanisms} describes mechanisms to preserve privacy of interactions, confidentiality of transactions and data, and confidentiality of business processes.
A guide for assessing DLT platforms with respect to their ability to meet specific enterprise requirements is provided in Section \ref{sec:guide}, that is subsequently applied to the use case of letters of credit in Section \ref{sec:usecase}.
Section \ref{sec:implementation} describes how privacy and confidentiality are addressed in the three largest DLTs for enterprise: Hyperledger Fabric (HLF) \cite{Androulaki2018}, Corda \cite{Hearn2016}, and Quorum \cite{JPMorganChase2016}.
The paper is concluded in Section \ref{sec:conclusion} with some final remarks.


\section{Privacy and Confidentiality Mechanisms}
\label{sec:mechanisms}
The following provides a reference to mechanisms that can be deployed to preserve privacy and confidentiality in DLT solutions, with a particular focus on their utility in enterprise solutions.
Privacy is used in the context of identity protection of individual parties as well as participants of a transaction.
Confidentiality is used when relating to the protection of data or business logic.
The available mechanisms to preserve privacy and confidentiality across solutions reach from structural design considerations to the use of cryptography.


\subsection{Privacy of interactions}
The verification of identities of parties onboarded to the platform is a common requirement of enterprise blockchain solutions.
This function is usually carried out by a service that allows parties to map public keys to identities through public key infrastructure (PKI) \cite{Rivest1978}.
This service may optionally expose a global membership list so that parties may establish relationships.
However, the group of parties entering into a business relationship often needs to be kept private from the wider network.

\paragraph{Separation of ledgers}
A network can be set up with not one global, but several private ledgers, where each is responsible for facilitating transactions between interacting parties.
This segregation can be established as a permanent network structure that holds a separate ledger as its own blockchain \cite{Androulaki2018}.
or on a per-transaction basis \cite{Hearn2016}, where data are sent only to involved parties.

\paragraph{One-time public keys}
In DLT platforms where ownership of assets is recorded against an address derived from a public key, one-time public keys can be used to mask the identity of the asset owner \cite{Hearn2016}.
Transacting parties and any entity that needs to verify signatures are then provided with a certificate that links the pseudonymous public key with an identity.

\paragraph{Zero-knowledge proof (ZKP) of identity}
ZKP is a cryptographic technique that can be used to prove that a party has a particular piece of information, or that a certain condition is met, without revealing the information itself.
In the context of identity, ZKPs can be used to prove the possession of a set of credentials without exposing the identity \cite{Camenisch2004}.
Using ZKPs, digital signatures from a party can be completely unlinkable to each other and to an identity.

\subsection{Confidentiality of transactions and data}
\label{transactions}
Transaction data are at the center of most business interactions.
Such data can contain trade secrets, financial records, or otherwise private agreements that may be sensitive not only to businesses but also to their clients.

\paragraph{Separation of ledgers}
Similarly to the way the separation of ledgers can provide privacy for interacting parties, it also provides confidentiality of data, revealing transaction data only to parties within the network partition.
If a public record of the existence of a transaction is required, a hash of transaction data may optionally be published on a shared ledger.

\paragraph{Off-chain data}
In cases where data need to be kept confidential to a subset of participants within a ledger, private data can be kept in an off-chain database.
This can either be natively integrated and hosted on a peer (peer off-chain), or be kept separate from the DLT layer entirely.
Transactions on the ledger can contain a hash of the off-chain data to provide authoritative evidence and an accompanying audit trail for involved parties to verify the provenance of private data.
Storing data off-chain has the additional property of enabling data to be deleted, for example, if required by law \cite{EuropeanCourtofJustice2014}.
However, allowing data deletion is in some way contradictory to the promise of an immutable, auditable record.

\paragraph{Symmetric key encryption}
Symmetric key encryption, for example the Advanced Encryption Standard (AES) \cite{NIST2001}, can be used to keep data confidential by encrypting values with a shared key between parties, which commonly gets shared over the network using PKI.
In order to implement this mechanism, use case and geographical region need to permit the sharing of encrypted data.

\paragraph{Merkle tree tear-offs}
One way of eliminating the need to share all data with all counterparties is to use a method called Merkle tree tear-offs \cite{Merkle1982}.
A Merkle tree is a data structure where every leaf is a hash of data, and every non-leaf node is a hash of the combined hashes of its child nodes.
In some DLTs, parties that are required to sign a transaction do so on the root of the Merkle tree constructed from all the transaction contents.
If some data within the transaction needs to be kept confidential from a party, the root of the confidential branch can be provided to them.
The party is able to compute and sign on the Merkle root without having access to the confidential data.

\paragraph{Multiparty computation}
Multiparty computation (MPC) \cite{Chaum1988} describes a collection of cryptographic algorithms that allows a group of parties to compute a shared function on private values.
Each party carries out a computation on their private data and shares the result with the other parties.
All collected results are then used by each party to compute the same shared function, resulting in one consistent value that can be committed to the ledger.
In DLT, this means that no private values need to be shared between parties and each participant would be able to store their data off-chain.
All functions and algorithms performed on the data are known to all involved parties.

\paragraph{Zero-knowledge proofs}
In the context of data confidentiality, a ZKP can be used to only provide enough information to prove that a certain fact is true (e.g. "the party has the appropriate funds") without revealing raw values \cite{Goldwasser1989}.
In enterprise DLTs, this becomes relevant when a precondition needs to be met before a smart contract will authorize and carry out a certain transaction, but the party in question does not want to reveal information beyond a boolean affirmation.
ZKPs need to be implemented specifically for a scenario and are currently only available for very specific functions \cite{Morais2018}.

\paragraph{Homomorphic computation}
Homomorphic encryption \cite{Gentry2010} describes cryptographic methods that allow for the computation of certain functions on encrypted input parameters to produce an equally encrypted output.
In theory this means that any party can carry out the computation on input data, vouch for correct execution, and thus be part of validation of transactions without being able to inspect any raw values.
Since homomorphic computation is still in the proof-of-concept stage and, moreover, has only been shown to enable a very limited set of operations, this method is infeasible for implementation in current systems \cite{Lauter2011}.

\paragraph{Trusted execution environments}
Trusted execution environments (TEEs) are hardware security modules within a CPU that guarantee confidentiality of executable code and data inside it \cite{Anati2013}.
Programs inside TEEs are physically isolated from the rest of the CPU, meaning no other software can access or modify them.
Each TEE owns a set of private keys that are embedded in the chip during manufacturing, with the corresponding public keys held by the manufacturer.
The TEE can provide an attestation of its state and the code running inside it, that can be signed by its private key, and is verifiable by the public key.
These features of TEEs mean they can be used to run smart contract code in a way that will keep both the code itself and the data around the smart contracts confidential.
Larger platforms have only recently started experimenting with TEEs \cite{Brandenburger2017} \cite{Limited2019}.

\subsection{Confidentiality of business logic}
A smart contract defines the conditions that need to be met when submitting a transaction to a ledger.
This can include a list of parties that need to endorse or sign a transaction as well as pre-defined logic that must be run to compute a valid parameter value to be committed.
Smart contract code needs to be distributed to parties that are required to endorse ledger updates to enable them to verify independently that proposed transactions abide by the agreed-upon logic.

\paragraph{Installation of smart contracts on involved nodes only}
Since smart contracts operate on a given ledger, the separation of ledgers means a separation of contracts, too, making them available only to members of the sub-network.
Regardless of how a network is organized however, it is desirable for a DLT to be able to distribute smart contracts only to those nodes that are needed for endorsement of transactions.

\paragraph{Off-chain execution engine}
Another mechanism to prevent business logic being revealed to non-involved parties, is through use of an off-chain execution engine \cite{Amundson}.
The smart contract code then only contains functions to read from and write to the ledger.
This not only prevents leaks of business logic, but also means the implementation is not bound to any particular programming language.
Additional challenges to enforce simultaneous updates across all engines for a particular ledger may arise.
Furthermore, some DLTs enforce participants to come to agreement on the smart contract before they can be used for transaction endorsement.

\paragraph{Trusted execution environments}
TEEs, as previously described in Section \ref{transactions}, provide an environment for the secure execution of code.
In DLTs, they can be used to execute smart contracts without allowing access to the clear text version of the logic.

\section{Design Guide}
\label{sec:guide}
Use cases and solutions are multifaceted.
Apart from use case driven privacy and confidentiality requirements, an architect may need to consider legal and regulatory constraints.
Furthermore, requirements may vary between different types of data.
This could mean that a solution needs to allow for personal data to be deleted (e.g. as per the General Data Protection Regulation (GDPR) \cite{EuropeanCourtofJustice2014}), or that encrypted data cannot be shared.
As such, an architect may choose to implement more than one method if different sets of data require different levels of confidentiality.

\subsection{Designing for privacy of interactions}
Platforms offer privacy control for different levels of granularity that can be tailored to the privacy requirements of a use case.
If a group of parties know each other, and members wish to interact privately, they may want to use a ledger that is  separate from the main chain.
If on any given ledger a sub-group of parties does not want to reveal that they are transacting they can exchange one-time public keys that cannot be linked directly to an identity.
In the case where an individual party wishes to remain entirely private but is still required to sign or commit a transaction, they have the ability to use ZKP to prove their identity.

\subsection{Designing for confidentiality of transactions}
\begin{figure}[!htb]
  \centering

  \tikzset{every picture/.style={line width=0.75pt}} 

  \begin{tikzpicture}[x=0.75pt,y=0.75pt,yscale=-1,xscale=1]

  \draw [color={rgb, 255:red, 65; green, 117; blue, 5 }  ,draw opacity=1 ]   (74,221) -- (106.17,235.19) ;
  \draw [shift={(108,236)}, rotate = 203.81] [fill={rgb, 255:red, 65; green, 117; blue, 5 }  ,fill opacity=1 ][line width=0.75]  [draw opacity=0] (8.93,-4.29) -- (0,0) -- (8.93,4.29) -- cycle    ;

  \draw (44.5,19) node [scale=0.7] [align=left] {Is data \\confidential?};
  \draw (44.5,152) node [scale=0.7] [align=left] {Can encrypted\\data be shared\\and stored?};
  \draw (140.5,151) node [scale=0.7] [align=left] {Data private to\\owner only?};
  \draw (219.5,153) node [scale=0.7] [align=left] {Boolean\\proofs\\enough?};
  \draw (43.5,68) node [scale=0.7] [align=left] {Is deletion\\necessary?};
  \draw (140.5,201) node [scale=0.7] [align=left] {Parts of data private\\to one or more parties?};
  \draw    (112,11) -- (181,11) -- (181,27) -- (112,27) -- cycle  ;
  \draw (146.5,19) node [scale=0.7] [align=left] {Single ledger};
  \draw    (191,228.5) -- (255,228.5) -- (255,280.5) -- (191,280.5) -- cycle  ;
  \draw (223,254.5) node [scale=0.7] [align=left] {Merkle tree\\tear-offs on \\separate \\ledger};
  \draw    (193,1.5) -- (246,1.5) -- (246,41.5) -- (193,41.5) -- cycle  ;
  \draw (219.5,21.5) node [scale=0.7] [align=left] {MPC with\\off-chain\\data};
  \draw    (205.5,194) -- (233.5,194) -- (233.5,210) -- (205.5,210) -- cycle  ;
  \draw (219.5,202) node [scale=0.7] [align=left] {ZKP};
  \draw  [color={rgb, 255:red, 128; green, 128; blue, 128 }  ,draw opacity=1 ][dash pattern={on 4.5pt off 4.5pt}]  (8.5,281.5) -- (81.5,281.5) -- (81.5,309.5) -- (8.5,309.5) -- cycle  ;
  \draw (45,295.5) node [scale=0.7] [align=left] {\textcolor[rgb]{0.5,0.5,0.5}{Homomorphic}\\\textcolor[rgb]{0.5,0.5,0.5}{computation}};
  \draw    (111.5,85) -- (172.5,85) -- (172.5,125) -- (111.5,125) -- cycle  ;
  \draw (142,105) node [scale=0.7] [align=left] {Off-chain \\data with\\public hash};
  \draw    (106,235) -- (176,235) -- (176,275) -- (106,275) -- cycle  ;
  \draw (141,255) node [scale=0.7] [align=left] {Separation of\\ledgers with\\optional hash};
  \draw (226.5,179) node [scale=0.5] [align=left] {Y};
  \draw (55.5,273) node [scale=0.5] [align=left] {N};
  \draw (54.5,231) node [scale=0.5] [align=left] {N};
  \draw (51.5,100) node [scale=0.5] [align=left] {N};
  \draw (94.5,14) node [scale=0.5] [align=left] {N};
  \draw (189.5,89) node [scale=0.5] [align=left] {N};
  \draw (227.5,54) node [scale=0.5] [align=left] {Y};
  \draw (53.5,38) node [scale=0.5] [align=left] {Y};
  \draw (126.5,64) node [scale=0.5] [align=left] {Y};
  \draw (148.5,220) node [scale=0.5] [align=left] {N};
  \draw (179.5,220) node [scale=0.5] [align=left] {Y};
  \draw (219.5,69) node [scale=0.7] [align=left] {Collective\\computation?};
  \draw    (106,283) -- (134,283) -- (134,299) -- (106,299) -- cycle  ;
  \draw (120,291) node [scale=0.7] [align=left] {TEE};
  \draw (44,206) node [scale=0.7] [align=left] {Are validators\\allowed to read \\transactions?};
  \draw (181.5,146) node [scale=0.5] [align=left] {Y};
  \draw (90.5,269) node [scale=0.5] [align=left] {\textbf{Y}/N};
  \draw (44,253) node [scale=0.7] [align=left] {Need to hide\\business logic?};
  \draw (92.5,224) node [scale=0.5] [align=left] {Y};
  \draw (53.5,178) node [scale=0.5] [align=left] {Y};
  \draw (87.5,145) node [scale=0.5] [align=left] {N};
  \draw (225.5,108) node [scale=0.5] [align=left] {N};
  \draw (149.5,171) node [scale=0.5] [align=left] {N};
  \draw [color={rgb, 255:red, 208; green, 2; blue, 27 }  ,draw opacity=1 ]   (83.5,151.59) -- (100.5,151.42) ;
  \draw [shift={(102.5,151.4)}, rotate = 539.4] [fill={rgb, 255:red, 208; green, 2; blue, 27 }  ,fill opacity=1 ][line width=0.75]  [draw opacity=0] (8.93,-4.29) -- (0,0) -- (8.93,4.29) -- cycle    ;

  \draw [color={rgb, 255:red, 208; green, 2; blue, 27 }  ,draw opacity=1 ]   (78,19) -- (110,19) ;
  \draw [shift={(112,19)}, rotate = 180] [fill={rgb, 255:red, 208; green, 2; blue, 27 }  ,fill opacity=1 ][line width=0.75]  [draw opacity=0] (8.93,-4.29) -- (0,0) -- (8.93,4.29) -- cycle    ;

  \draw [color={rgb, 255:red, 65; green, 117; blue, 5 }  ,draw opacity=1 ]   (219.5,173) -- (219.5,192) ;
  \draw [shift={(219.5,194)}, rotate = 270] [fill={rgb, 255:red, 65; green, 117; blue, 5 }  ,fill opacity=1 ][line width=0.75]  [draw opacity=0] (8.93,-4.29) -- (0,0) -- (8.93,4.29) -- cycle    ;

  \draw [color={rgb, 255:red, 65; green, 117; blue, 5 }  ,draw opacity=1 ]   (178.5,151.96) -- (193,152.33) ;
  \draw [shift={(195,152.38)}, rotate = 181.45] [fill={rgb, 255:red, 65; green, 117; blue, 5 }  ,fill opacity=1 ][line width=0.75]  [draw opacity=0] (8.93,-4.29) -- (0,0) -- (8.93,4.29) -- cycle    ;

  \draw [color={rgb, 255:red, 65; green, 117; blue, 5 }  ,draw opacity=1 ]   (44.21,33) -- (43.83,52) ;
  \draw [shift={(43.79,54)}, rotate = 271.17] [fill={rgb, 255:red, 65; green, 117; blue, 5 }  ,fill opacity=1 ][line width=0.75]  [draw opacity=0] (8.93,-4.29) -- (0,0) -- (8.93,4.29) -- cycle    ;

  \draw [color={rgb, 255:red, 208; green, 2; blue, 27 }  ,draw opacity=1 ]   (43.67,82) -- (44.24,130) ;
  \draw [shift={(44.26,132)}, rotate = 269.32] [fill={rgb, 255:red, 208; green, 2; blue, 27 }  ,fill opacity=1 ][line width=0.75]  [draw opacity=0] (8.93,-4.29) -- (0,0) -- (8.93,4.29) -- cycle    ;

  \draw [color={rgb, 255:red, 208; green, 2; blue, 27 }  ,draw opacity=1 ]   (219.5,133) -- (219.5,85) ;
  \draw [shift={(219.5,83)}, rotate = 450] [fill={rgb, 255:red, 208; green, 2; blue, 27 }  ,fill opacity=1 ][line width=0.75]  [draw opacity=0] (8.93,-4.29) -- (0,0) -- (8.93,4.29) -- cycle    ;

  \draw [color={rgb, 255:red, 65; green, 117; blue, 5 }  ,draw opacity=1 ]   (219.5,55) -- (219.5,43.5) ;
  \draw [shift={(219.5,41.5)}, rotate = 450] [fill={rgb, 255:red, 65; green, 117; blue, 5 }  ,fill opacity=1 ][line width=0.75]  [draw opacity=0] (8.93,-4.29) -- (0,0) -- (8.93,4.29) -- cycle    ;

  \draw [color={rgb, 255:red, 208; green, 2; blue, 27 }  ,draw opacity=1 ]   (189.36,83) -- (174.31,89.99) ;
  \draw [shift={(172.5,90.83)}, rotate = 335.08000000000004] [fill={rgb, 255:red, 208; green, 2; blue, 27 }  ,fill opacity=1 ][line width=0.75]  [draw opacity=0] (8.93,-4.29) -- (0,0) -- (8.93,4.29) -- cycle    ;

  \draw [color={rgb, 255:red, 65; green, 117; blue, 5 }  ,draw opacity=1 ]   (44.31,172) -- (44.2,184) ;
  \draw [shift={(44.19,186)}, rotate = 270.53] [fill={rgb, 255:red, 65; green, 117; blue, 5 }  ,fill opacity=1 ][line width=0.75]  [draw opacity=0] (8.93,-4.29) -- (0,0) -- (8.93,4.29) -- cycle    ;

  \draw [color={rgb, 255:red, 65; green, 117; blue, 5 }  ,draw opacity=1 ]   (72,267) -- (104.21,283.11) ;
  \draw [shift={(106,284)}, rotate = 206.57] [fill={rgb, 255:red, 65; green, 117; blue, 5 }  ,fill opacity=1 ][line width=0.75]  [draw opacity=0] (8.93,-4.29) -- (0,0) -- (8.93,4.29) -- cycle    ;

  \draw [color={rgb, 255:red, 208; green, 2; blue, 27 }  ,draw opacity=1 ]   (44,226) -- (44,237) ;
  \draw [shift={(44,239)}, rotate = 270] [fill={rgb, 255:red, 208; green, 2; blue, 27 }  ,fill opacity=1 ][line width=0.75]  [draw opacity=0] (8.93,-4.29) -- (0,0) -- (8.93,4.29) -- cycle    ;

  \draw [color={rgb, 255:red, 208; green, 2; blue, 27 }  ,draw opacity=0.5 ] [dash pattern={on 4.5pt off 4.5pt}]  (44.33,267) -- (44.62,279.5) ;
  \draw [shift={(44.67,281.5)}, rotate = 268.65] [fill={rgb, 255:red, 208; green, 2; blue, 27 }  ,fill opacity=0.5 ][line width=0.75]  [draw opacity=0] (8.93,-4.29) -- (0,0) -- (8.93,4.29) -- cycle    ;

  \draw [color={rgb, 255:red, 208; green, 2; blue, 27 }  ,draw opacity=1 ]   (140.5,165) -- (140.5,185) ;
  \draw [shift={(140.5,187)}, rotate = 270] [fill={rgb, 255:red, 208; green, 2; blue, 27 }  ,fill opacity=1 ][line width=0.75]  [draw opacity=0] (8.93,-4.29) -- (0,0) -- (8.93,4.29) -- cycle    ;

  \draw [color={rgb, 255:red, 208; green, 2; blue, 27 }  ,draw opacity=1 ]   (140.63,215) -- (140.8,233) ;
  \draw [shift={(140.81,235)}, rotate = 269.47] [fill={rgb, 255:red, 208; green, 2; blue, 27 }  ,fill opacity=1 ][line width=0.75]  [draw opacity=0] (8.93,-4.29) -- (0,0) -- (8.93,4.29) -- cycle    ;

  \draw [color={rgb, 255:red, 65; green, 117; blue, 5 }  ,draw opacity=1 ]   (162.09,215) -- (189.32,232.66) ;
  \draw [shift={(191,233.75)}, rotate = 212.96] [fill={rgb, 255:red, 65; green, 117; blue, 5 }  ,fill opacity=1 ][line width=0.75]  [draw opacity=0] (8.93,-4.29) -- (0,0) -- (8.93,4.29) -- cycle    ;

  \draw [color={rgb, 255:red, 65; green, 117; blue, 5 }  ,draw opacity=1 ]   (74.5,68.18) -- (182,68.79) ;
  \draw [shift={(184,68.8)}, rotate = 180.33] [fill={rgb, 255:red, 65; green, 117; blue, 5 }  ,fill opacity=1 ][line width=0.75]  [draw opacity=0] (8.93,-4.29) -- (0,0) -- (8.93,4.29) -- cycle    ;

  \end{tikzpicture}

  \caption{Guide to mapping confidentiality requirements on data to available techniques.}
  \label{tab:data}
\end{figure}
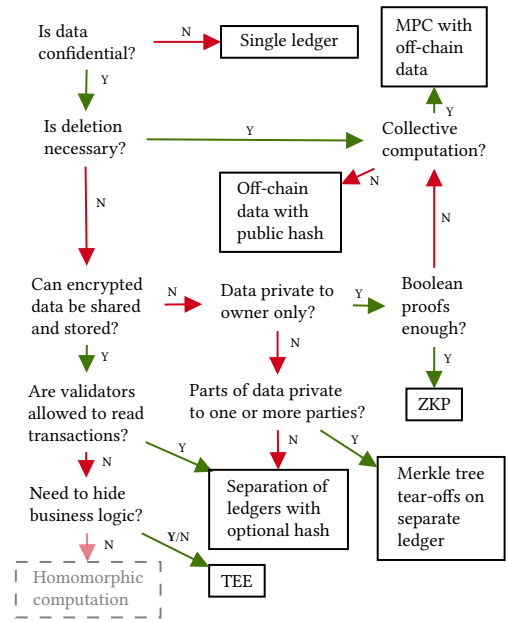

Figure \ref{tab:data} aims to guide the reader in mapping transaction confidentiality requirements to available mechanisms.
A first important decision point involves regulatory obligations, such as "the right to be forgotten" \cite{EuropeanCourtofJustice2014}.
Since distributed ledgers inherently do not allow for the removal of entries, data need to be kept off-chain if deletion is required.
Note that some ledger implementations offer the ability to "prune" the chain to allow archiving of older transactions \cite{LinuxFoundation}, \cite{Hearn2016}.
However, archived entries are generally still available to parties on request.
Given enough computing resources, encrypted data can be decrypted, which means that parties may prefer not to share even encrypted data with the wider network.
If on-chain records are still desired to make use of endorsement protocols or the append-only character of a ledger, this will usually lead to the implementation of segregated ledgers with constrained membership.
Additional Merkle tree tear-offs can be implemented if a transaction contains data irrelevant to one or more participating parties and must be kept private.
Unless uninvolved network parties are required to endorse the correctness of an otherwise confidential transaction, segregated ledgers may more generally be the preferred solution.
Note that by storing a hash of data on a shared ledger, it is recorded that a transaction occurred without revealing its content.
If independent validation while keeping data confidential is desirable, uninvolved nodes can provision trusted execution environments, which provides the added benefit that business logic need not be revealed.
Homomorphic computation, while not mature enough to date, may also eventually enable the processing of encrypted values.
In some cases, a transaction may rely on private data that cannot be shared between transacting parties.
Zero-knowledge proofs can provide boolean affirmation, for example to prove that a party has sufficient funds.
If a shared function needs to be computed on private values, such as a would be the case for a secret ballot, multiparty computation can be used.
Most enterprise platforms are putting continued effort into advancing ZKPs and MPC to make them natively available.
Not captured in this diagram is the case where a node is administered by a third party that may not be trusted with raw data.
In that case, transaction data can be encrypted through symmetric or asymmetric cryptography.

\subsection{Designing for confidentiality of business logic}
There are a number of factors that may influence the choice of mechanism to keep business logic private.
Four criteria an architect may want to consider are whether an implementation (1) keeps logic private, (2) offers in-built smart contract versioning, (3) hides data from the node administrator, and/or (4) allows for business logic to be written in any programming language.
There may be network configurations in which a node is administered by a third party that should not have access to unencrypted data or business logic.
For the case where contract code requires access to the confidential encrypted data, it is possible to run computations in a trusted execution environment.
If this level of confidentiality is not needed but business logic remains confidential, contracts can be installed only on involved nodes or alternatively can be run using a separate execution engine.
A separate engine allows for the free choice of programming language, which may be especially relevant for sectors that use domain-specific languages.
However, an external engine will not benefit from the mechanism in-built to most DLT platforms that ensures that all nodes run the same version of smart contract code, meaning that version control will need to be managed outside the DLT layer.

\subsection{Common technical challenges}
\label{challenges}

Enterprise solutions offer a variety of techniques to ensure that parties, data, and code are kept private using network design and encryption methods.
However, when permissioning a platform, some other factors need to be considered that are independent of the particular DLT being used.

\paragraph{Ordering transactions}
On a distributed ledger ensuring all nodes in the network agree on the same state (or at least the part of the state they are entitled to see) is imperative.
The service that provides ordering of transactions to construct a correct view of the state is an integral part of any DLT platform.
For some of the platforms reviewed (Fabric and Corda), this service has visibility of all DLT events, including parties to transactions and transaction details.
When assessing a DLT for suitability, architects must consider whether the ordering service meets privacy and confidentiality requirements and if parties can feasibly run their own service to mitigate leaks.

\paragraph{Permissioning infrastructure}
Ultimately, any DLT solution will need to be hosted on some organization's infrastructure, potentially exposing ledger entries and transactions to system administrators.
To prevent any leak, ideally, a network should be designed such that all layers of the application (i.e. user interface, middleware, DLT) can be hosted on a per-organization level, giving each party on the network the ability to fully control their own environment.
Parties should optionally be able to run a node on the cloud, choosing from a number of Blockchain-as-a-Service providers.
When choosing a DLT platform for enterprise, it is useful to be aware of which cloud providers natively support the chosen solution.
In cases where businesses cannot or do not wish to manage their own infrastructure, they may need to rely on an external provider, trusting a third party with maintaining privacy and confidentiality.
In this way, financial and time constraints may require an organization to compromise on privacy and/or confidentiality.

\paragraph{Performance and scalability}
The benefits of a DLT solution become more obvious the more parties are sharing one business network.
This inherently means that at some point, performance at scale of the solution will need to be assessed.
While enterprise platforms commonly offer one core mechanism to protect private and confidential information, there are still unanswered questions around how these solutions scale.
This is partly due to the lack of clarity on which metrics should be used in the context of permissioned DLT, and partly due to inherent difficulties in comparing the different approaches consistently.
Scalability of confidentiality preserving methods on HLF, Corda, and Quorum have partially been addressed in \cite{Ferris2019}, \cite{Hearn2016}, and \cite{Baliga2018}, respectively.
However, when designing a solution, custom scalability tests may need to be designed to fit the particular use case.

\section{Example Use Case}
\label{sec:usecase}
A letter of credit is a financial instrument in which a bank vouches to pay a seller if a buyer is unable to make an agreed-upon payment.
Parties on a DLT network used to record letters of credit are banks, sellers, and buyers.
Sellers and buyers will neither want to share that they are entering in a business relationship nor the details of their agreement with the network.
Under the assumption that logic contained in a letter of credit is highly standardized and non-confidential, the design guide will lead to the following design.

Identities of parties will need to be verified by an independent party, most likely a bank.
Since, according to GDPR regulations, any party is allowed to request deletion of personally identifiable information, according to Figure \ref{tab:data} any such data will need to be stored off-ledger.
All non-personal data will not be required to be deleted and can therefore be included in transactions.
We will work under the assumption that there is no regulation against the sharing and storing of encrypted data.
The solution can then be designed such that transaction validators will be the parties associated with the transaction, and therefore will have the authority to view the transaction contents.
According to Figure \ref{tab:data}, these requirements lead to a design where groups of transacting parties use a separate ledger in order to keep their interactions hidden from the wider network.
If a third party is trusted to run the ordering service and have visibility of transacting parties, transaction data can be encrypted.


\section{Enterprise Implementations}
\label{sec:implementation}
\begin{center}
\begin{table}
\begin{tabular}{c L{5cm}|c c c c c c}
     & Mechanism & \rot{HLF} & \rot{Corda} & \rot{Quorum} \\
 \specialrule{2pt}{1pt}{1pt}
 \multirow{3}{*}{\rot{Parties}}
 &Separation of ledgers                                         & \tick      & \tick     & \tick     \\
 &One-time public key                                              & \dash      & \tick     & $\star$ \\
 &Zero knowledge proof of identity                              & \tick      & \dash     & \dash     \\
 \hline
 \multirow{7}{*}{\rot{Transactions}}
 &Separation of ledgers                                         & \tick      & \tick     & \tick     \\
 &Off-chain peer data                                                & \tick      & $\star$   & \dash    \\
 &Symmetric keys                                                  & \tick      & \tick     & \tick     \\
 &Merkle trees and tear-offs                                      & $\star$    & \tick     & \dash     \\
 &Zero-knowledge proofs$^1$                                       & $\star$    & $\star$   & $\star$   \\
 &Multiparty computation                                      & $\star$    & $\star$   & $\star$   \\
 &Homomorphic encryption$^1$                                      & $\star$    & $\star$   & $\star$ \\
 \hline
 \multirow{3}{*}{\rot{Logic}}
 &Install contract on involved nodes                            & \tick      & N/A    & \tick     \\
 &Off-chain execution engine                                     & $\star$    & \tick    & \dash     \\
 &Trusted execution environments$^1$                             & \dash      & \dash    & \dash      \\
 \hline
 \multirow{2}{*}{\rot{Misc.}}
 &Private sequencing service possible                          & \tick       & \tick    & \tick    \\
 &Open source                                                  & \tick       & \tick    & \tick    \\

\end{tabular}
\caption{
Comparison of permissioned DLTs with respect to privacy and confidentiality mechanisms. \tick: native support, $\star$: not natively supported, but can be implemented, \dash: requires substantial rewriting of the code base. $^1$ Maturity level described in Section \ref{sec:mechanisms}.
}
\label{tab:comparison}

\end{table}
\end{center}

This section describes how the privacy and confidentiality mechanisms discussed in Section \ref{sec:mechanisms} have been implemented in three of the most prominent permissioned blockchain platforms - Hyperledger Fabric, Corda, and Quorum.
These DLTs were chosen because they are a) open-source, b) backed by large and active communities, and c) represent three different ways in which a certain level of privacy and confidentiality can be achieved.
An overview of natively supported methods and extendibility of these platforms is given in Table \ref{tab:comparison}.
This section provides a point-in-time evaluation only.
The teams behind included platforms are actively researching ways to improve privacy and confidentiality further, predominantly in the cryptography domain.
In the coming years, it is therefore likely that there will be more native implementations of advanced cryptographic techniques such as ZKPs, homomorphic encryption, MPC, and TEEs.

\paragraph{Hyperledger Fabric}
Fabric is an open source permissioned blockchain under the umbrella of the Linux Foundation's Hyperledger project.
It was designed specifically for enterprise use and has a strong focus on privacy and confidentiality.
The primary mechanisms for privacy and confidentiality preservation is through channels, which provide a separate ledger for a subset of participants \cite{Androulaki2018a}.
Identities of channel members are not revealed to the wider network and transactions are only shared between channel members.
Confidentiality of smart contract logic, called chaincode, is provided by ensuring only peers that have the chaincode installed are able to view the chaincode.
The exception to the confidentiality boundary of the channel is the service used to provide consensus on the order of transactions.
In Fabric, the ordering service has full visibility of channel members as well as all transactions that are submitted to a channel.
This is a potential breach of privacy and confidentiality and is discussed in greater detail in Section \ref{challenges}.
In order to mitigate the risk of leaking information, channel members can choose to run the ordering service themselves given appropriate infrastructure.
Within a channel, Fabric provides privacy of parties with Idemix \cite{Camenisch2004}, enabling zero-knowledge proof of identity using the public key of the issuing certificate authority to verify the credentials rather than disclosing the identity.
Confidential data is also possible between sub-groups of channel participants through Private Data Collections (PDCs), which allow for data to be kept off the channel ledger (off-chain) and referenced in transactions by hash only.
However, members of PDCs are listed in associated transactions, so this method of confidentiality preservation is useful only if privacy of interaction is not required within the channel.

\paragraph{Corda}
Corda, an open source DLT, is being developed by R3, a consortium consisting of mostly financial institutions.
Rather than globally broadcasting transactions to all peers in the network or a sub-network, Corda uses a concept of peer-to-peer transactions.
The peer-to-peer nature of Corda naturally lends itself to confidentiality of data through segregation; interactions between parties are kept private, both in terms of the relationships that exist and data shared between them.
If assets are to be transferred beyond the initial transacting parties, Corda allows the use of one-time public keys in transactions to further conceal identities from uninvolved parties.
For situations where parties within a single transaction should not have access to all transaction data, Corda also provides support for Merkle tree tear-offs.
A common scenario for this is when an oracle is needed to attest to a certain piece of data in a transaction, but the transaction participants do not want all the components of the transaction visible to the oracle.
Corda provides confidentiality of business logic by separating the execution of business logic from the verification of valid transactions.
The rules that define which parties are required to sign a transaction are contained in contracts associated with each state that is provided as input to a transaction.
These parties execute business logic outside of the platform to determine whether the transaction proposal is valid, giving parties the added freedom to choose a programming language to implement business logic.
The on-chain contract is used to verify the transaction's signatories.

\paragraph{Quorum}
JPMorgan partnered with the Ethereum Enterprise Alliance to develop Quorum, a permissioned blockchain based on the Ethereum protocol.
Its key differentiator is the ability to store private state separate from the public ledger.
This separation of private state from public state is the primary mechanism for confidentiality of data and smart contracts.
Private state and smart contracts are updated through private transactions that are distributed to all nodes in the network.
However only a hash of the submitted data is included in the transaction itself.
The parties involved in the transaction receive encrypted data, which means decryption is required before a party can update their private state.
One key limitation of the private transaction model in Quorum is that it does not prevent the double spending of assets.
The contents of private transactions are shared only between specified parties.
Since there is no global visibility of private assets, a party may spend an asset multiple times by specifying different receivers in separate transactions.
Another major drawback of Quorum is that the public ledger includes private transactions, including the list of participants of the transaction, revealing to the entire network which parties are interacting.

\section{Conclusion}
\label{sec:conclusion}
\label{discussion}
From transparent execution of smart contracts to increased data integrity and availability, DLT has the potential to provide a new framework for the way collaborative business processes are conducted.
However, business interactions can contain highly sensitive information, including personal data, and a leak thereof could have large economic impact.
Any enterprise use case may require that privacy of interacting entities be preserved and that both the data and the business logic that define how data are updated and kept confidential.
In DLT implementations, these requirements will likely determine how the network is structured and what privacy and confidentiality mechanisms can be deployed.
This paper lays out relevant key mechanisms for meeting said requirements, covering segmentation and encryption techniques.
It further gives the reader a guide to both define their particular needs and discover possible solutions.
To illustrate how the mechanisms are applied in current DLTs, this paper described how three representative DLTs for enterprise implement privacy and confidentiality.
Apart from native support in existing enterprise solutions, financial restrictions, scalability of a solution, and infrastructure administration may also play an important role in realizing the ideal solution.

\textbf{Acknowledgements} The authors thank reviewers of this paper, namely E. Abebe, Y. Hu, D. Karunamoorthy, E. Lloyd, L. Mehedy, C. Vecchiola, J. Wagner, and N. Waywood.


\bibliography{bibliography}
\bibliographystyle{ACM-Reference-Format}

\end{document}